# Stabilization of the bias point in MZM modulators

Zhuo Wang[1]

[1]*School of Optics and Photonics, Beijing Institute of Technology*

**Abstract:** This article mainly introduces the role of MZM in practical communication systems, the materials used to make MZM modulators such as lithium niobate, and its working principle. It also explains why it changes due to environmental factors. This leads to the introduction of a method that controls the stable points of MZM by algorithmically controlling the voltage, and the algorithm is verified through experiments. Finally, a summary and outlook on the future development of MZM are provided.

## 1. Research background and significance

Over the past few years, research on integrated optical devices based on lithium niobate materials, such as integrated optical electro-optic modulators, optical switches, electric field sensors, and magnetic field sensors, has attracted the attention of many researchers. However, these integrated optical devices require an appropriate bias operating point in practical work, that is, applying the appropriate bias phase to the device to make it work properly[1]. For example, lithium niobate external modulators applied in fiber optic communication systems also need a stable bias phase, so that the bit error rate of the optical transmission system can be effectively reduced; when lithium niobate is applied to the field of optical switches, such optical switch devices also need to be biased at the appropriate operating point, so that the transmittance of the optical switch can flexibly switch between the maximum and minimum values, and at this time the corresponding devices are fixed at the "0" and "π" operating points, which is conducive to improving the extinction ratio of the device; fiber optic gyroscopes made based on the Sagnac interference principle also include a modulator with a Y-waveguide structure, and fiber optic gyroscopes are precision devices, so they have extremely high requirements for the stability of the modulator's operating point. However, in the practical engineering applications of lithium niobate devices, some internal and external factors such as temperature, external electric fields, stress, etc., as well as defects in the device itself, will have a very significant impact on the stability of the modulation phase of lithium niobate integrated optical devices, causing the operating point of the device to drift. The amplitude of the drift can sometimes be small and sometimes very large, and the frequency of the drift can sometimes be small and sometimes large. Under the influence of these factors, the integrated optical devices cause the operating point to be unstable. When the amplitude and frequency of the drift are too large, it will cause the lithium niobate integrated optical device to fail, which limits its application in engineering[2].

### 1.1 Lithium niobate-based MZM modulator

Y-shaped waveguide modulators, intensity modulators, phase modulators, optical switches, and electric field sensors used in fiber optic gyroscopes are all major components of lithium niobate optical waveguides[2].

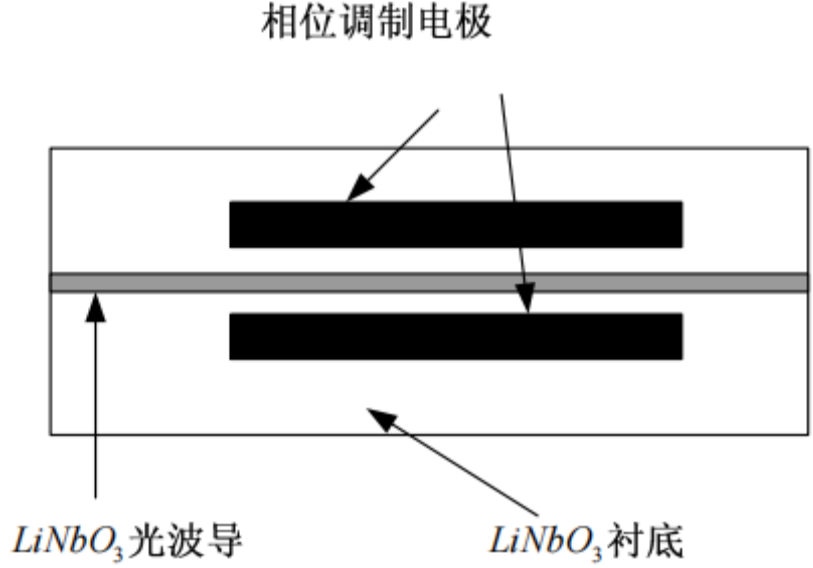

Fig. 1. A Schematic diagram of the phase-wise modulator[2]

Figure 1 shows the basic structure of a lithium niobate phase modulator. The phase modulator, combined with an optical waveguide, can form other types of modulators. The linear electro-optic effect of lithium niobate material is the fundamental working principle of these devices.
Assuming the propagation direction of the optical waveguide is along the z-axis and it has good ability to confine light, and it is a single-mode optical waveguide. Assuming its refractive index distribution is n(x,y), according to the linear electro-optic effect of lithium niobate, the refractive index distribution will change under the influence of the applied electric field Es:

$$\Delta n(x,y) = -\frac{1}{2}n^3 r E_s \qquad (1\text{-}1)$$

which causes a change in the effective refractive index of the waveguide:

$$\Delta n_{eff} = -\frac{1}{2}n^3 r \langle E_s, |E_0|^2 \rangle = -\frac{1}{2}n^3 r \frac{\Gamma}{g} V \qquad (1\text{-}2)$$

The corresponding phase shift can be seen in equation (1-3), thus forming the modulation of the applied electric field on the phase of the guided mode.

$$\Delta \phi = \Delta \beta L = \frac{2\pi}{\lambda_0}\Delta n_{eff} L = -\frac{\pi}{\lambda_0} n^3 r \frac{\Gamma}{g} V L \qquad (1\text{-}3)$$

Figure 2 is a schematic diagram of a structure using lithium niobate as the substrate. It utilizes the coupling of the evanescent fields of two closely spaced optical waveguides. By changing the propagation phase of the two waveguides, it achieves the switching of the waveguide between the two waveguides. Under the drive of the switching voltage, the optical switch has a stable on or off state, which also requires stable phase modulation performance or switching operating points.
Under the influence of many factors such as temperature, stress (including thermal stress and mechanical stress, etc.), and applied voltage, the actual electric field applied to the waveguide changes. Sometimes these factors can even change the effective refractive index of the waveguide. Thus, the modulation phase change of the integrated optical modulator under a fixed voltage will cause the bias operating point to drift. Therefore, it is of significant importance to analyze the bias point drift of the integrated optical modulator in detail and to take certain measures to stabilize the bias phase of the lithium niobate integrated optical device.

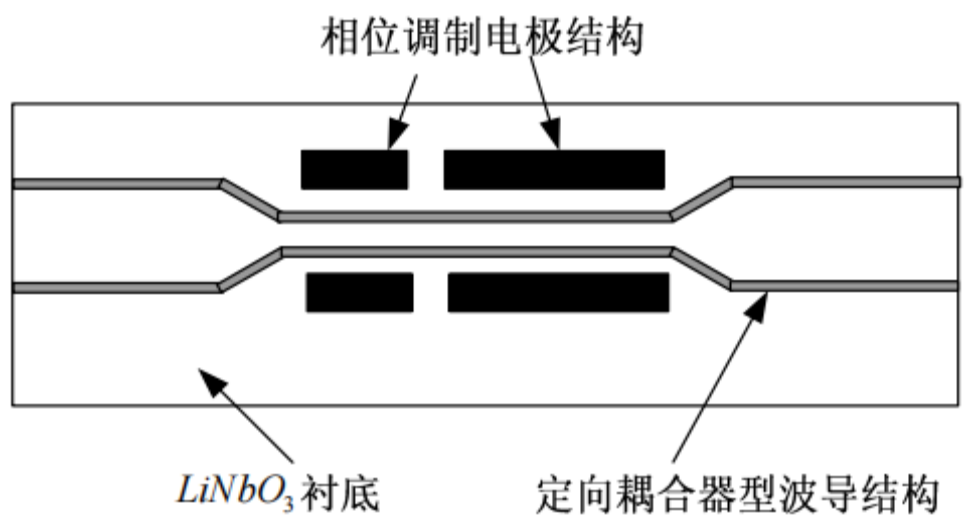

Fig. 2. Principle diagram of a directional coupler type optical switch[2]

*1.2 The principle behind the drift of the operating point in lithium niobate.*

(1) The DC drift principle of lithium niobate waveguide modulators

As the basic structural unit of lithium niobate waveguide modulators, the cross-sectional view of the phase modulator is shown in Figure 3. The silica layer has two functions in the device. One is to reduce the absorption loss of the waveguide by the electrodes when they cross over the surface of the waveguide, while minimizing the impact of the electrodes on the waveguide's mode field distribution as much as possible[4]. The second is to effectively adjust the microwave effective refractive index of the modulator, so as to achieve the matching of the waveguide mode and the propagation speed of the modulation signal, thus making the lithium niobate modulator a high-speed modulation device. However, if the silica layer is not deposited on the lithium niobate modulator, then the electro-optic effect of the modulator will be stronger, which can significantly reduce the modulator's voltage; at the same time, the amplitude of the DC drift in such devices without the silica layer is also reduced. This type of X-cut lithium niobate integrated optical device without the silica layer can operate normally for 20 years at room temperature of 650°C, and at that time, the amplitude of the DC drift will not even be greater than the original voltage applied to the modulator. Although devices without the silica layer can effectively suppress the DC drift of the device, the structure of most lithium niobate integrated optical devices uses a structure with a silica layer. Therefore, the improvement of the DC drift phenomenon needs to be achieved by optimizing other structures of the device.

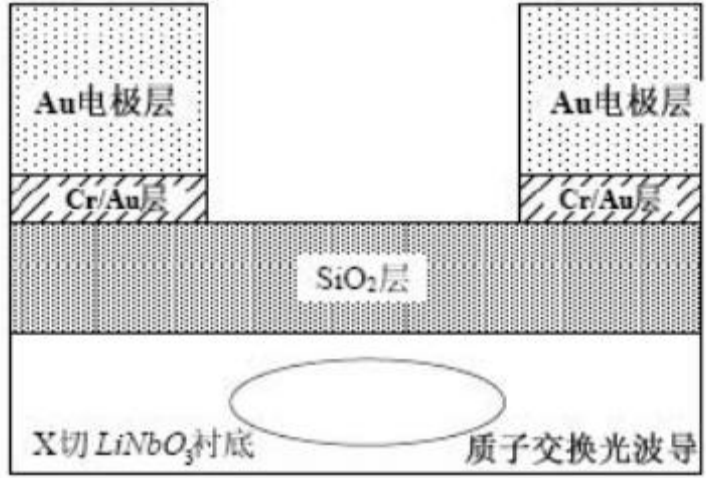

Fig. 3. Cross-sectional view of the basic unit structure of a lithium niobate waveguide modulator[4]

The model of the equivalent circuit diagram as shown in Figure 4 is often used to analyze the DC drift characteristics of lithium niobate modulators. In the diagram, the resistances of the buffer layer in the transverse and longitudinal directions are represented by $R_1$, $C_1$ and $R_2$, $C_2$, respectively. $R_3$ and $C_3$ represent the resistance and capacitance of the lithium niobate waveguide, respectively, and Ei represents the magnitude of the electric field formed by space charge.

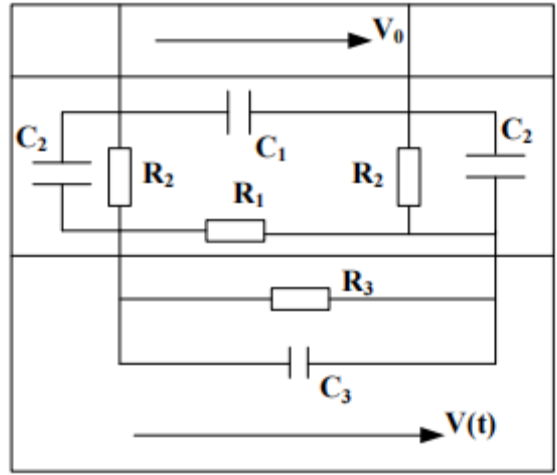

Fig. 4. Equivalent circuit diagram of the basic unit structure of lithium niobate waveguide modulator[4]

Assuming the step signal voltage value applied to the electrodes is V0, the voltage applied to the waveguide can be represented as:

$$V(t) = V_0 \left[ \frac{R_4}{2R_2 + R_4} + \frac{2(C_2 R_2 - C_3 R_4)}{(2R_2 + R_4)(C_2 + 2C_3)} e^{-t/\tau} \right] \quad (1-4)$$

In the above equation, $R_4$ represents the effective lateral impedance, and $\tau$ is the relaxation time, with the following: $R_4 = R_1 R_3/(R_1+R_3)$, $\tau = R_2 R_4 (C_2+2C_3)/(2R_2+R_4)$。

When the relaxation time is much smaller than the bias time of the signal, the effective voltage value applied to the waveguide is:

$$V(t \gg \tau) = V_0 R_4 / (2R_2 + R_4) \qquad (1\text{-}5)$$

From formulas (1-4) and (1-5), it can be seen that the effective voltage value applied to the lithium niobate modulator waveguide varies with time, which is the phenomenon known as the modulator's DC drift. The main reason for this phenomenon is primarily due to the electrical characteristics of the substrate material, waveguide, buffer layer, and other media. However, the electrical properties between the various medium layers are also an important cause of this phenomenon. Therefore, controlling the electrical performance of the substrate material and buffer layer (such as dielectric constant, conductivity, etc.) can effectively suppress the DC drift of lithium niobate modulation devices.

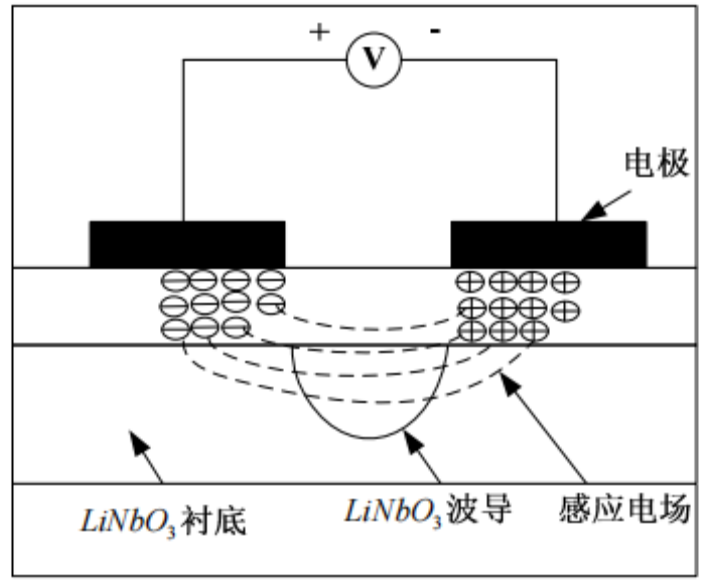

Fig. 5. The spatial electric field formed by movable charges under the action of an external electric field

The movement of charges present in the substrate and buffer layers under the influence of an applied electric field can be considered one of the reasons for the occurrence of DC drift in devices. Figure 5 illustrates the movement of charges within the modulator under the influence of an applied electric field. Free ions such as H+, K+, Na+ exist in the substrate and buffer layers within the device structure, which move to the sides of the waveguide, thereby creating a spatial electric field across the waveguide that is opposite in direction to the electric field formed on the waveguide due to the externally applied DC voltage on the electrodes. As a result, the effective electric field acting on the waveguide is weakened. The lithium niobate substrate also contains metal impurities such as Fe, which can be excited to the conduction band by the light transmitted through the waveguide, thus generating positive charges.

$$Fe^{2+} + h\nu \rightarrow Fe^{3+} + e^{-1} \qquad (1\text{-}6)$$

When no external electric field is applied, other energy levels capture these excited electrons from the conduction band. An internal electric field is formed between these energy levels and the original impurity levels, which causes a change in the refractive index of the lithium niobate crystal through its electro-optic effect. The DC drift caused by this reason is usually a few seconds to minutes, sometimes even longer. The content of hydrogen ions has a significant impact on the stability of the operating point of the device.

(2) The effect of temperature on the operating point of lithium niobate crystals

The fabrication of lithium niobate optical waveguides often employs processes such as titanium diffusion or proton exchange[5]. When the propagation direction of the waveguide is in the z-direction, the distribution of the refractive index of the crystal in the xy-direction is:

$$n^2(x,y) = \begin{cases} n_b^2 + (n_s^2 - n_b^2)e^{-(x/d_x)^2 \cdot (y/d_y)^2}, & y \geq 0 \\ n_c^2, & y \leq 0 \end{cases} \qquad (1\text{-}7)$$

In the above formula, nc represents the refractive index value of the waveguide cladding, nb is the refractive index value of the substrate, ns is the peak refractive index of the waveguide surface, and dx, dy denote the waveguide thickness in the x and y directions, respectively, as shown in Figure 6.

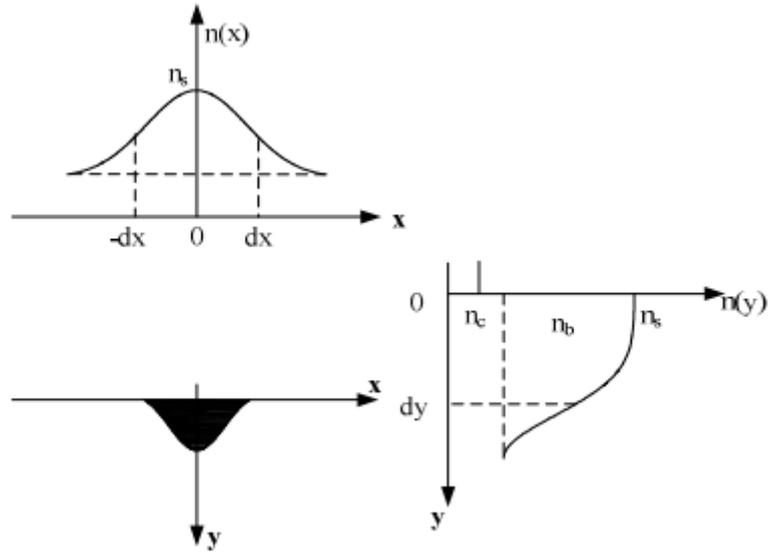

Fig. 6. Schematic diagram of refractive index distribution in lithium niobate waveguide[5]

The normalized frequency of a waveguide is defined as:

$$\begin{cases} v_x = k_0 d_x \sqrt{n_s^2 - n_b^2} \\ v_y = k_0 d_y \sqrt{n_s^2 - n_b^2} \end{cases} \qquad (1\text{-}8)$$

The normalized propagation constant is defined as:

$$(n^2 - n_b^2) / (n_s^2 - n_b^2) \qquad (1\text{-}9)$$

In the above equation, n represents the effective refractive index of the waveguide. Since the refractive index of lithium niobate material is a function of wavelength and temperature, the relationship between the change in refractive index and temperature can be expressed by equation (1-10).

$$\frac{1}{n_e}\frac{\partial n_e}{\partial T} = 17.1\times10^{-6}\,K^{-1},$$
$$\frac{1}{n_0}\frac{\partial n_0}{\partial T} = 1.9\times10^{-6}\,K^{-1} \qquad (1\text{-}10)$$

In practical single-mode waveguides, the normalized frequency is approximately around 3.0, and the waveguide's form ratio is between 1.5 to 2.5. Such a waveguide structure can achieve strong mode confinement and minimize the waveguide's modal radiation loss to the greatest extent[6]. The rate of change of the waveguide normalized frequency with temperature (δV)T is about $10^{-5}$ $K^{-1}$, and the rate of change of the waveguide's effective refractive index with temperature (δn)T is roughly $10^{-6}$ $K^{-1}$. Therefore, temperature generally has a relatively small impact on the performance of the waveguide itself. To avoid the phenomenon of lithium niobate modulator's working point drift, various methods can be adopted, such as selecting appropriate materials to reduce the content of various impurity ions in the device structure, which can effectively suppress the device's working point drift characteristics. Moreover, mechanical methods and specific device fabrication process optimizations can also effectively inhibit the lithium niobate modulator's working point drift. However, these methods make the device fabrication process complex and difficult to implement. Currently, the most commonly used method is to employ an additional feedback system to automatically control the device's bias phase. The feedback signal required for the feedback control system can be derived not only from the ratio of the input and output DC optical power of the modulator but also from the amplitude values of the fundamental and second harmonic signals in the low-frequency jitter signal modulated onto the optical path.

## 2. A stable scheme for MZM modulator based on a composite control algorithm using the average optical power slope value and cotangent value.

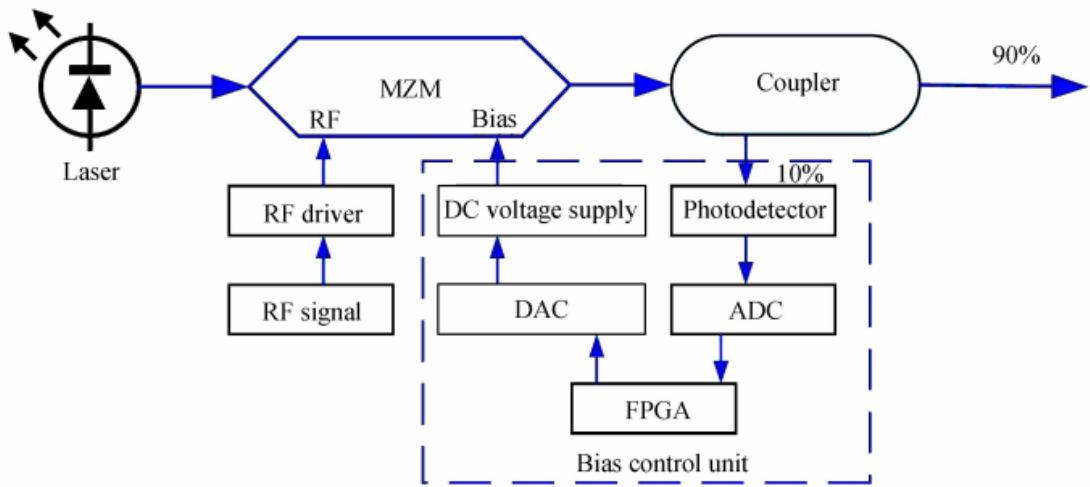

Fig. 7. Control diagram of the bias operating point for a Mach-Zehnder modulator[7]

The designed MZM bias operating point control system is shown in Figure 7. A modulation drive signal is inputted to the modulator pulse input end (RF), and the corresponding bias voltage is determined according to the modulation format[7]. The optical signal outputted by the MZM enters the bias control unit through a 9:1 optical coupler. The photodetector converts the optical signal into an electrical signal, which is then processed by the FPGA for AD and DA conversion. When $V_{Bias} \neq mV_\pi$ (m=1,3,⋯), the FPGA reads the output voltage value $V_{g1}$ of the photodetector. After increasing the bias voltage ΔV at the MZM's Bias end, the FPGA reads the output voltage value $V_{g2}$ of the photodetector again. The average optical power slope value d11 is calculated based on the two collected voltage values. Then, the bias voltage ΔV is increased again at the MZM's Bias end, and the average optical power slope value $d_{13}$ is calculated. The second derivative of the average optical power $d_2$ is calculated from $d_{11}$ and $d_{13}$, and the initial cotangent value R1 is calculated from d11 and $d_2$. Repeat these steps to obtain $R_2$; when $R_2 > R_1$, it indicates that the operating point has shifted to the left. The FPGA outputs a compensation voltage according to the range of the interval, which is then converted and amplified through DA conversion and inputted to the MZM bias voltage input end until $R_1 = R_2$ (the compensation method is opposite when $R_2 < R_1$). When $V_{Bias} = mV_\pi$ (m=1,3,⋯), the FPGA only needs to calculate the slope value of the average optical power. By comparing the size relationship between this slope value $d_{12}$ and 0, it is determined whether the bias operating point has drifted. If the operating point has drifted, adjust the bias voltage until $d_{12} = 0$.

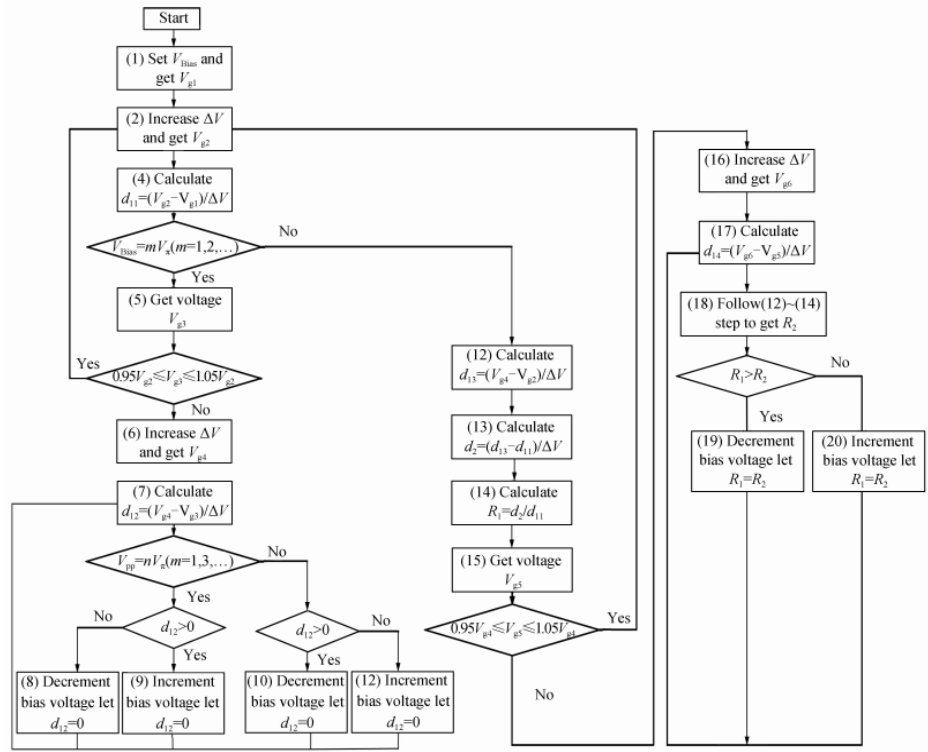

Fig. 8. Mach-Zehnder Modulator Bias Control Flowchart[7]

The bias working point control flowchart is shown in Figure 8. Here, $V_{g1}$ is the voltage value initially collected by the FPGA after setting the bias voltage $V_{Bias}$, $V_{g2}$ is the voltage value collected at the output end after applying a voltage of $V_{Bias}$+ΔV to the input end of the electro-optic modulator, $V_{g3}$ is the voltage value collected at the output end after calculating the average optical power slope value and when $V_{Bias}=mV_\pi$ (m=1,3,...), Vg4 is the voltage value collected at the output end after applying a voltage of $V_{Bias}$+2ΔV to the input end of the electro-optic modulator, $V_{g5}$ is the voltage value collected after calculating the initial cotangent value R1, and Vg6 is the voltage value collected at the output end after applying $V_{Bias}$+ΔV again to the input end of the electro-optic modulator (if the transmission response curve of the electro-optic modulator changes at this point, the collected voltage value corresponding to $V_{Bias}+\Delta V$ on the new transmission response curve is different from the collected voltage value corresponding to the original transmission response curve, that is: $V_2 \neq V_6$).

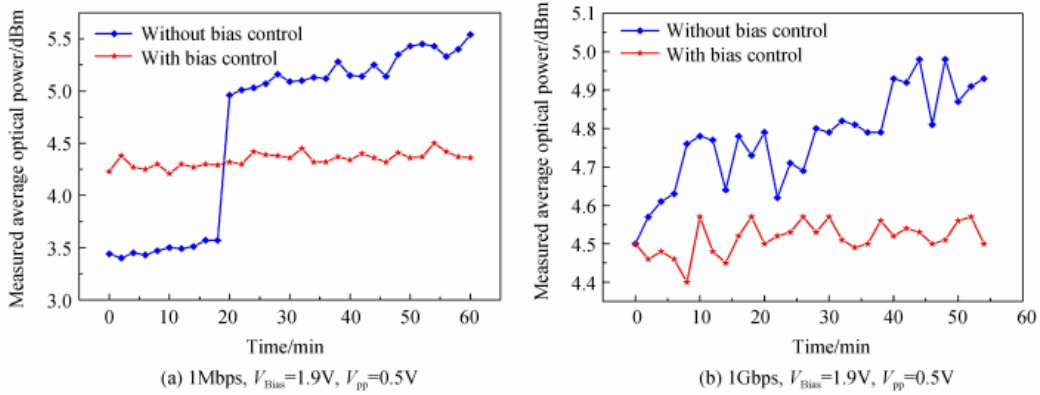

Fig.9 The test result of AOP for working bias control system[7]

Firstly, the change in the average optical power output of the MZM was tested without using the bias point control at a communication rate of 1Mbps. As shown by the dotted line in Figure 8(a), the bias point continuously drifted and the output average optical power gradually changed over one hour. At the 20th minute, the output average optical power suddenly changed from 3.57dBm to 4.96dBm. During the test period, the output average optical power changed from the initial 3.44dBm to the final 5.14dBm, with a fluctuation of 1.7dBm. After using the bias point control system, as indicated by the triangular line in Figure 9(a), the bias point drifted slightly within one hour, and the fluctuation of the output average optical power was 0.21dBm. At a communication rate of 1Gbps, without using the bias point control, the output average optical power changed from 4.5dBm to 4.93dBm. As shown by the diamond line in Figure 9(b), within one hour, the output average optical power had two significant fluctuation ranges, which were 4.6dBm~4.75dBm and 4.8dBm~5.0dBm, with a fluctuation of 0.43dBm within one hour. As shown by the triangular line in Figure 9(b), after using the bias point control, the fluctuation of the output average optical power was only 0.24dBm. From the above test data, it is evident that, at different communication rates, using the bias point control results in better stability of the system's output average optical power.

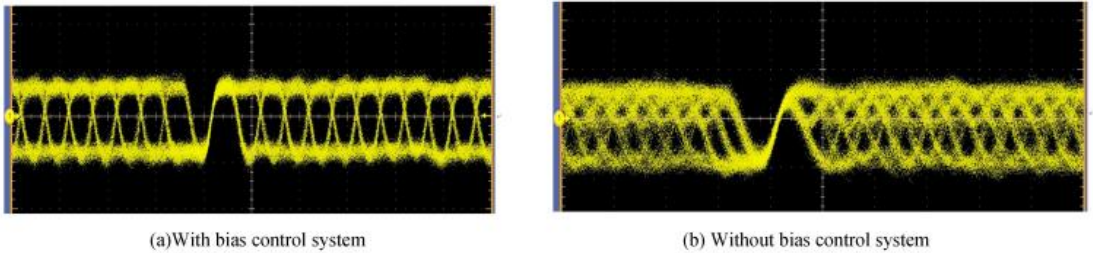

Fig.10 The eye diagram of working bias control system with 10 Gbps[7]

When the bias voltage of the MZM is 1.9V, the bias operating point is located at the quadrature point of the transmission response curve. At this point, OOK modulation is performed, and the output optical signal power of the MZM is approximately 4.5dBm. The eye diagram of the modulation signal at a communication rate of 1Gbps is shown in Figure 10. Figure 10(a) shows the eye diagram using the bias operating point control. As can be seen, the eye opening is moderate, inter-symbol interference is small; the trace of the eye diagram is clear, and the noise interference is low. The eye diagram without using bias operating point control is shown in Figure 10(b). As can be seen, the eye opening is small, the trace is blurred, the noise interference is large, the eye diagram is not well-formed, and the bit error rate is high. From the system test eye diagram, it can be seen that using bias operating point control can improve the phenomenon of communication quality deterioration caused by the drift of the bias operating point.

## 3. Summary and Outlook

This article analyzes and discusses the impact of the drift of the bias operating point of a Mach-Zehnder modulator on the quality of optical communication[8-10]. Based on this, a composite bias point control scheme based on the average optical power slope value and cotangent value is introduced. Experimental results show that the bias operating point control system can keep the fluctuation of the output average optical power within ±5% at both 1Mbps and 1Gbps communication rates, and the eye diagram quality is good. Compared with the pilot harmonic control method and the direct optical power control method, this scheme has the advantages of simple system structure, good control accuracy, small size, and strong reliability. The control scheme can stabilize the bias operating point at any position on the transmission response curve and can be widely applied to externally modulated optical communication systems that perform intensity and phase modulation.